\documentstyle[preprint,aps]{revtex}
\begin{document}
\draft
\title{A new twist on $Z \rightarrow b {\bar b}$}
\author{Pham Quang Hung}
\address{Department of Physics, University of Virginia, Charlottesville,
Virginia 22901, USA}
\date{\today}
\maketitle
\begin{abstract}
A new mechanism is proposed to explain the "anomaly" in
$Z \rightarrow b {\bar b}$ resulting in a prediction of a new,
{\em non-sequential} fourth family whose masses could all
be below $M_W$, thus opening up an exciting prospect for
near-future discoveries at LEP2 and possibly at the Tevatron.
\end{abstract}
\pacs{12.15.-y, 12.15.Lk, 12.60.Rc, 13.38.Dg}

\narrowtext

Precision tests of the Standard Model (SM) have reached a level where it
"might" now be possible to look for indirect evidence of new physics and/or
new degrees of freedom. One example  is the {\em apparent} discrepancy
between theory and experiment in the value of the ratio
$R_{b} \equiv \Gamma (Z \rightarrow b {\bar b})/
\Gamma (Z \rightarrow had )$. This discrepancy which increases
with $m_t$, reaches the 2 $\sigma$ level when $m_t$ reaches 175 GeV.
If one also includes the {\em apparent} disagreement between the
QCD coupling $\alpha_{S}$ determined at "low" energy and evolved to
$M_Z$ with that determined by the Z-lineshape, one is tempted to
think that one might be already seeing some new kind of physics.
It is therefore very crucial
to confirm or disprove these so-called discrepancies.
Let us nevertheless assume that they are there to stay
and examine what kind of new physics that can be possible and
what predictions that can be tested in the near future. (Even if
the discrepancy were to disappear, this would put a severe constraint
on this type of new physics.)

In this letter, a mechanism is proposed to explain the apparent
increase of $R_b$ and to make further predictions
on other branching ratios, and ultimately on
the new physics concerning the mechanism itself.
This mechanism is based on the assumption
that there is a new, heavy, {\em non-sequential} down
quark ($Q = -1/3$) ( part of a new family) with mass greater than 46 GeV
and whose $q {\bar q}$ bound state(s) mixes with the
Z boson. By {\em non-sequential},
we mean that the fermions of
the new family does not have ( or has little)
mass mixing with fermions of the other
three generations. Other than this being a working assumption,
a realization of this scenario is given at the end of the paper.

For this paper,
we shall quote a few relevant observables \cite{langacker,glasgow}:
$\Gamma_Z (GeV) = 2.4974 \pm 0.0027 \pm 0.0027 (2.496 \pm 0.001 \pm 0.003
\pm [0.003])$,
$R_b \equiv \Gamma(b {\bar b}) / \Gamma(had) = 0.2202 \pm 0.0020
(0.2155 \pm 0 \pm 0.0004)$,
$R_c \equiv \Gamma(c {\bar c}) / \Gamma(had) = 0.1583 \pm 0.0098
(0.171 \pm 0 \pm 0)$, where the numbers in parentheses are the
standard model expectations, and
$R_e \equiv \Gamma(had) / \Gamma(e {\bar e}) =
20.850 \pm 0.067$, $R_{\mu} \equiv \Gamma(had) /
\Gamma(\mu {\bar \mu}) = 20.824 \pm 0.059$.


Let us denote this {\em non-sequential} family by ($\cal R$,
$\cal P$) for the quarks and by ($\cal N$, $\cal E$) for
the leptons. For reasons to be given below let us assume
that the ($Q=-1/3$) quark has a mass $m_{\cal P} < m_{\cal R}$.
We also assume that the up-type quark ${\cal R}$ is heavy
enough so that ${\cal R}{\bar{\cal R}}$ bound states are
well above the ${\cal P}{\bar{\cal P}}$ open threshold. 
The ${\cal P} {\bar{\cal P}}$ bound states can be described
by Richardson's potential. Such an analysis has been carried out
long ago by \cite{gilman} for the $ ^3S_1$ $t {\bar t}$
bound states, but unfortunately in the now-obsolete range of
$m_t \sim 40-50$ GeV. This analysis can however be applied to
any quark in a similar mass range or higher, especially
for our case where $m_{{\cal P}} > 46$ GeV. 
(The mass shift of the Z boson due to this mixing is negligible
\cite{gilman}.)

${\cal P} {\bar {\cal P}}$ bound states which can mix with Z
are either vector, axial vector, or both. In what follows
we shall neglect the mixing of Z with the axial vector states
since it goes like $\beta^3$ \cite{zerwas,gilman}. Consequently
we shall focus only on the vector meson ($^3S_1$) bound states.
In particular, we shall first examine the mixing of the ground
state $1S$ with Z. In the mass range considered here, the ground
state $1S$ is sufficiently far from open-$\cal P$ threshold so
that  the mass-mixing formalism can be applied. Denoting the
$1S$ ($J^{PC} = 1^{--}$) state by $V^0$, the result of $V^0$
and $Z^0$ mixing is given in terms of the mass eigenstates \cite{gilman}
\begin{mathletters}
\begin{equation}
|V\rangle = cos\frac{\theta}{2} |V_0\rangle - sin\frac{\theta}{2}
|Z_0\rangle,
\label{massvec1}
\end{equation}
\begin{equation}
|Z\rangle = sin\frac{\theta}{2} |V_0\rangle + cos\frac{\theta}{2}
|Z_0\rangle,
\label{massvec2}
\end{equation}
\end{mathletters}
for the mass eigenvectors and where
\begin{equation}
\theta = sin^{-1} (\delta m^2 / \Delta^2),
\label{angle1}
\end{equation}
with
$\Delta^2 = [\frac{(
M_{V_0}^2 - i \Gamma_{V_0} M_{V_0} - M_{Z_0}^2 + i \Gamma_{Z_0} M_{Z_0})^2}
{4}+ (\delta m^2)^2 ]^{1/2}$.
$\delta m^2$ is the off-diagonal element of the mass mixing matrix
and is given by \cite{gilman}
\begin{equation}
\delta m^2 = F_V [(\frac{g}{cos\theta_W})\frac{\frac{4}{3}sin^2\theta_W
-1}{4}],
\label{offdiag}
\end{equation}
where
$F_V = 2 \sqrt{3} |\Psi (0)| \sqrt{M_{V_0}}$.
The term inside the square brackets represents the {\em vector}
coupling of the $\cal P$ quark to the Z boson.

Let us assume that $M_V > M_Z$ and since present experiments are carried
out on the Z resonance, we need only to look at Eq.\ (\ref{massvec2})
to see how the presence of $V_0$ modifies the coupling of Z to
"light" quarks and leptons. This, as we claim in this manuscript,
is a possible source for the discrepancy seen in $\Gamma(b {\bar b})$.
{}From Eq.\ (\ref{massvec2}), one finds the physical Z couplings to
a given fermion $f$ to be
\begin{equation}
g_{Z f {\bar f}}^{V,A} = sin\frac{\theta}{2} g_{V_0 f {\bar f}}^{V,A}
+ cos\frac{\theta}{2} g_{Z_0 f {\bar f}}^{V,A},
\label{coupling}
\end{equation}
where $V$ and $A$ stand for vector and axial-vector couplings respectively.

The most obvious source of the coupling of $V_0$ to $f {\bar f}$ is via
$\gamma$ and Z and evaluated at $s = M_Z^2$.
The electroweak source alone however gives only a small change to $R_b$ and
in the wrong direction, and this worsens when $V_0$ is
close in mass to Z. A new and unconventional coupling
of $\cal P$ to $b$ ( and to other normal fermions as well)
is needed, not only to compensate for this small electroweak
change but also to bring $R_b$ closer to its experimental
value. To this end, let us write
\begin{equation}
g_{V_0 f {\bar f}}^{V,A} = F_V G_{f}^{V,A}(s=M_Z^2) +
g_{new, f}^{V,A},
\label{coupling2}
\end{equation}
where \cite{barger}
\begin{mathletters}
\begin{equation}
G_{f}^{V}(M_Z^2) = e^2 \frac{Q_f Q_{\cal P}}{M_Z^2} +
\frac{g^2}{cos^2\theta_W} \frac{g_f^V g_{\cal P}^V}{M_Z^2 -
M_{Z_0}^2 + i M_{Z_0}\Gamma_{Z_0}},
\label{GV}
\end{equation}
\begin{equation}
G_{f}^{A}(M_Z^2) = \frac{g^2}{cos^2\theta_W}
\frac{g_f^A g_{\cal P}^A}{M_Z^2 - M_{Z_0}^2 + i M_{Z_0}\Gamma_{Z_0}},
\label{GA}
\end{equation}
\end{mathletters}
and where $g_f^{V,A}$ is the vector (axial-vector) coupling of the
Z boson to the fermion $f$, and $g_{\cal P}^V = - (1-(4/3)sin^2
\theta_W)/4$ and $g_{\cal P}^A = 1/4$. $Q_f$ and $Q_{\cal P}( = -1/3)$
are the electric charges. $g_{new,f}^{V,A}$ is the coupling of
$V_0$ to a fermion $f$ coming from  some new physics.
Since $M_Z^2 \simeq M_{Z_0}^2$ and
$M_{Z_0}\Gamma_{Z_0} \simeq 3 \times 10^{-2} M_Z^2$, we can
safely neglect the $\gamma$ contribution in Eq.\ (\ref{GV})
(it contributes negligibly to the present discussion).

For the mass range considered below (shown in the Figure),
namely $m_{\cal P} \simeq 48 GeV-53 GeV$, $|\Psi (0)|$ is
such that \cite{gilman} $|\delta m^2| \ll |
M_{V_0}^2 - i \Gamma_{V_0} M_{V_0} - M_{Z_0}^2 +
i \Gamma_{Z_0} M_{Z_0})^2|/2$ and consequently
\begin{equation}
sin\frac{\theta}{2}\approx \frac{\delta m^2}{M_{V_0}^2
- M_{Z_0}^2 + i (\Gamma_{Z_0} M_{Z_0} - \Gamma_{V_0} M_{V_0})},
\label{sint}
\end{equation}
with $cos\frac{\theta}{2} \approx 1$.
Typically, $\theta/2 \approx 2-3 \times 10^{-2}$ and the deviation
of $cos\frac{\theta}{2}$ from unity will be of order $10^{-4}$ and
can neglected considering the present level of precision.

The modified couplings of Z to a fermion $f$ are now
\begin{mathletters}
\begin{equation}
\tilde{g}_{f}^V = (1+ \eta_{f,W}^V + \eta_{f,new}^V) g_{f}^V,
\label{gvtil}
\end{equation}
\begin{equation}
\tilde{g}_{f}^A = (1+ \eta_{f,W}^A + \eta_{f,new}^A) g_{f}^A,
\label{gatil}
\end{equation}
\end{mathletters}
where $W$ stands for electroweak and the $\eta$'s are complex numbers
and are defined by
\begin{mathletters}
\begin{equation}
\eta_{f,W}^{V,A}  = sin\frac{\theta}{2} F_V G_{f}^{V,A}(s=M_Z^2)/g_{f}^{V,A},
\label{etawre}
\end{equation}
\begin{equation}
\eta_{f,new}^{V,A}  = sin\frac{\theta}{2} g_{new}^{V,A}/g_{f}^{V,A},
\label{etanre}
\end{equation}
\end{mathletters}
where the explicit forms for $\eta_{f,W}^{V,A}$ and $\eta_{f,new}^{V,A}$
can be obtained by using Eqs.\ (\ref{coupling2},\ref{GV},\ref{GA},\ref{sint}).
{\em For simplicity}, we shall now assume that $g_{new,f}^{V} = g_{new}
 \neq 0$ and $g_{new,f}^{A} = 0$ so as to reduce the number of
parameters and to study the implication of such an assumption.
The introduction of a {\em single} new coupling is what
we meant by universality earlier. We shall argue below
why we expect such a behavior. (The inclusion of $g_{new,f}^{A}$ is
quite straightforward.)

In computing the Z widths using Eqs.\ (\ref{gvtil},\ref{gatil}) and the
range of mass mentioned earlier, one can safely {\em neglect}
terms proportional to $(Re\:\eta)^2$ and $(Im\:\eta)^2$ since
they turn out to be {\em at least} two orders of magnitude smaller
than terms proportional to $Re\:\eta$ (assuming $g_{new,f}^{V} < 1$).
(Considering the present level of precision, their inclusion is
irrelevant to the present discussion.) With this remark in mind,
the decay width for $Z \rightarrow f {\bar f}$ is now given by
\begin{equation}
\Gamma(Z \rightarrow f {\bar f}) = \Gamma_{f}^{SM} (1 +
\delta_{new}^{f}),
\label{gam}
\end{equation}
where $f = q, l$ and where
\begin{equation}
\delta_{new}^{f} = \frac{2 ((g_{f}^{V})^2 ( Re\:\eta_{W}^V +
Re\:\eta_{new}^{V}) + (g_{f}^{A})^2 Re\:\eta_{W}^{A})}{
(g_{f}^{V})^2 + (g_{f}^{A})^2}.
\label{delnew}
\end{equation}
In Eq.\ (\ref{gam}), $\Gamma_{f}^{SM}$ contains various radiative
correction factors as well as mass factors such as defined in Ref.\
(\cite{Bernabeu}). We find
\begin{mathletters}
\begin{eqnarray}
\Gamma(had)& = &\Gamma^{SM}(had) + \delta_{new}^{u} ( \Gamma_{u}^{SM}
+ \Gamma_{c}^{SM}) \nonumber \\
& & \mbox{}+ \delta_{new}^{d} (\Gamma_{d}^{SM}
+ \Gamma_{s}^{SM} + \Gamma_{b}^{SM}),
\label{gamhad}
\end{eqnarray}
\begin{equation}
R_{f} =  \frac{R_{f}^{SM}(1+ \delta_{new}^f)}{1 +
\delta_{new}^{u} ( R_{u}^{SM}
+ R_{c}^{SM}) + \delta_{new}^{d} (R_{d}^{SM}
+ R_{s}^{SM} + R_{b}^{SM})},
\label{rhad}
\end{equation}
\end{mathletters}
where $R_{f} \equiv \Gamma(Z \rightarrow q_f {\bar q}_f) / \Gamma(had)$.
The central theme of this paper is the use of $R_b$ to obtain
information on the model proposed here. By using Eq.\ (\ref{rhad})
for $R_b$, one can extract the parameter $Re\eta_{b,new}^{V}$ and
consequently the {\em common} parameter $sin\frac{\theta}{2}g_{new}$
as a function of $M_{V_0}$. This will then be used to make predictions
on various ratios mentioned above and also on the total Z width.
In particular, $R_{e,\mu}$ will be used to constrain the
range of allowed $M_{V_0}$.

We shall use the following hadron ratios: $R_b = 0.2202 \pm 0.0020$,
$R_{b}^{SM} = 0.2155$, $R_{c}^{SM} = 0.1721$,
$R_{s}^{SM} = 0.22$, $R_{u}^{SM} = 0.1722$. The Standard Model
predictions \cite{Bernabeu} given here are for $m_t = 170$ GeV.
A more extensive analysis using the range of $m_t$ given by
CDF and D0 will be given elsewhere.
Notice that the Standard Model values quoted
here are rather insensitive to the Higgs boson mass.

We {\em predict}: $R_c = 0.165 \mp 0.003$,
$R_s = 0.225 \pm 0.002$,
$R_u = 0.165 \mp 0.003$ to be compared with
$R_{c,exp} =0.1583 \pm 0.0098$ (more than 1$\sigma$ lower than
the standard model prediction).

Notice that
an increase in the ratio for a down-type quark corresponds to
a decrease in the ratio for an up-type quark and vice versa.
This happens because $Re\eta_{f,new}^{V}$ is positive for
$f=u,c$ and negative for $f=d,s,b$. ($V_0$ is a ${\cal P}
{\bar{\cal P}}$ bound state.)

Let us turn to $R_{e,\mu}$. The experimental values are
$R_e \equiv \Gamma(had) / \Gamma(e {\bar e}) =
20.850 \pm 0.067$, $R_{\mu} \equiv \Gamma(had) /
\Gamma(\mu {\bar \mu}) = 20.824 \pm 0.059$ to be compared
with the Standard
Model expectations ($m_t = 170$ GeV):
$R_{e}^{SM} = R_{\mu}^{SM} = 20.774, \ldots , 20.754$ for $m_H =
100 , \ldots , 1000 $ GeV. Although these numbers are
consistent within errors (except for $R_e$),
experimentally there seems to be
a tendancy for an increase in these ratios. (Even when we
take into account the spread in $m_t$, $R_{e,\mu}^{SM} <
20.78$.) Our predictions for $R_e = R_{\mu}$ are shown in
the Figure (curves labeled by 300 and 700)
as a function of $M_{V_0}$ and for two different
values of $m_H$, namely $m_H= 300, 700$ GeV.
They are obtained by using $R_{b,min}$ ( the theoretical curves
obtained by using $R_{b,max}$ are out of scale in the Figure
presented here). The predicted regions lie above
these curves. The two vertical
lines represent the {\em lower} bounds on $M_V$, namely
$M_{V_0} = 95.6, 103.25$ GeV for
$m_H = 300, 700$ respectively, coming from the $\Gamma_Z$ constraint. The
arrows indicate that the regions allowed by $\Gamma_Z$ are to the right.
Finally the two horizontal lines delimit the experimentally
allowed region which we take to be the overlap
between $R_e$ and $R_{\mu}$. There we take 20.883 as the maximum (
from $R_{\mu}$) and 20.783 as the minimum (from $R_e$).
We only show $m_H$
up to 700 GeV to be consistent with the global fit although
larger values are entirely possible. (For $m_H = 1000$ GeV,
the lower bound on $M_V$ is 95.2 GeV.)

The allowed regions are the ones bounded
by the theoretical curves, the vertical lines and $R_{e,exp}$.
{}From the Figure one can see that the allowed region for
$m_H = 700$ GeV is {\em significantly} larger than that for
$m_H = 300$ GeV. This implies that if the resonance were
to be found say at 96 GeV one would infer $m_H> 700$ GeV,
while if it were found at 103.6 GeV one would have
a looser bound, namely $m_H> 300$ GeV. A lower $M_{V_0}$
implies a higher lower bound on $m_H$. We can also
conclude that, for $m_H < 1000$ GeV, the resonance mass which
is {\em compatible} with {\em all} available data is
bounded from below by 95.2 GeV. This corresponds to
$m_{{\cal P}} \geq 48.5$ GeV.



{}From $R_b$, we can extract $g_{new}$ as a function
of $M_{V_0}$ and use these values to compute $\Gamma_{V_0}$.
The dominant contribution to $\Gamma_{V_0}$ turns out to
come mainly from this $g_{new}$ with $\gamma$, Z, and three
gluon processes contributing a small amount to the total
width. For $M_{V_0} > 95.2$ GeV, the lower bound on
$\Gamma_{V_0}$ is found to be 1 GeV and increasing to
approximately 8 GeV for $M_{V_0} \approx 104$ GeV. The
expectation is in general a few GeVs for the width
in our scenario while a standard heavy onium will have
a width of at most a few MeVs. Notice that
the shift in width due to the mixing with Z is small with
respect to the above intrinsic width. A further prediction
is the fact that, in our scenario, the new coupling is
universal so that $V_0$ couples equally to quarks and
leptons of both up and down types. This implies that
$\Gamma(V_0 \rightarrow l {\bar l}) = \Gamma_0$ and
$\Gamma(V_0 \rightarrow q {\bar q}) = 3 \Gamma_0$.
The prediction for the branching ratios is
$B_l = 1/24$ and $B_q = 1/8$ where $l= e,\mu,\tau,\nu_{e,\mu,\tau}$
and $q=u,d,s,c,b$.

Let us now turn to the other members of this {\em non-sequential}
family, the ${\cal R}$ quark and the leptons ${\cal N}$ and
${\cal E}$.
This is where the S and T parameters \cite{peskin} come in.
Since this new family is {\em non-sequential}, there is no reason
to expect the mass splitting between up and down members to be
"similar" to the other three families.
We use the results of \cite{langacker} (a seven parameter
fit) which show the allowed regions in the $T_{new}^{'}-S_{new}^{'}$
plane where $T_{new}^{'}= T_{new} + T_{M_H}$ and
$S_{new}^{'}= S_{new} + S_{M_H}$. In view of the discrepancy
between the SLD and the LEP asymmetries, we shall use as the
allowed region the overlap of all data except the SLD asymmetries.
In particular we would like to ask whether or not all of these
new particles can be lighter than 80 GeV, an exciting scenario
since they can all be accessible to LEP2 in that case. Since
the possibilities are many, a few examples will be illuminating.

Let us take $m_{{\cal P}} = 50$ GeV and $m_H = 700$ GeV. The Higgs
contribution to S and T is given by $S_{M_H} = 0.045$ and
$T_{M_H} = -0.132$. For $m_{{\cal R}} = 60, 80$ GeV, one
has $S_{new,quark} = 0.138, 0.107$ and $T_{new,quark} = 0.497, 0.686$
respectively. For the leptons, we shall assume \cite{gates}
that ${\cal N}$ is
a Majorana particle and that its mass ( as well as
that of ${\cal E}$) is greater than 46 GeV.
This is the only direct constraint one has on the leptons. We
shall use two representative sets of values.
1) For $m_{{\cal N}}=46$ GeV, $S_{new,lepton} = 0.055, 0.027$
and $T_{new,lepton} = -0.004, 0.011$ for
$m_{{\cal E}} = 60, 80$ GeV respectively. We then get
the following results in terms of ($S_{new}^{'}, T_{new}^{'}$)
for the pair ($m_{{\cal R}}, m_{{\cal E}}$). We
obtain: (0.238, 0.361) for (60, 60) GeV, (0.207, 0.55)
for (80, 60) GeV, (0.21, 0.376) for (60, 80) GeV, and
(0.179, 0.565) for (80, 80) GeV.
2) For $m_{{\cal N}} = 60$ GeV, $S_{new,lepton} = 0.073, 0.054$,
$T_{new,lepton} = -0.013, -0.007$ for
$m_{{\cal E}} = 65, 80$ respectively. We obtain:
(0.256, 0.352) for (60, 65) GeV, (0.225, 0.541) for (80, 65) GeV,
(0.237, 0.358) for (60, 80) GeV, and (0.206, 0.547) for
(80, 80) GeV. All of those values are inside the
allowed region shown in \cite{langacker}.

{}From the (not-exhaustive)examples given above, it is clear that
the S and T parameters certainly allow for the existence
of this new, {\em non-sequential} family and that the T parameter
tends to favor a value of ${\cal R}$ mass lower than 80
GeV (and in no way should it be more than 90 GeV). This opens up
the possibility that the whole family can be found by LEP2.
First the R ratio would be 16/3 or at least 14/3 (if the
${\cal R}$ quark mass is above 80 GeV). Secondly, there
would be {\em two} narrow resonances: the first one being
the ${\cal P}{\bar{\cal P}}$ bound state and the second one
being the ${\cal R}{\bar{\cal R}}$ bound state. Since these
quarks do not mix with with those of the other three
generations, ${\cal P}$ will be relatively stable and
the search for ${\cal R}$ will be an interesting problem
in hadron colliders. Finally there will be an unmistakable
signature for the charged heavy lepton in LEP2. These
issues will be explored elsewhere.

Finally we would like to say a few words about a possible origin
of the coupling $g_{new}$. First solving for $g_{new}$ using
Eq. \ (\ref{rhad}), one has for example the following range:
$g_{new,min} = 0.135-0.377$ for $M_{V_0} = 95.5-104.5$ GeV
where $R_{b,min}$ has been used. (It turns out that the
constraint coming from $R_{e,\mu}$ gives an allowed range
for $g_{new}$ ranging from the previous $g_{new,min}$ to a
slightly higher value for each $M_{V_0}$.) It is clear that
the larger $M_{V_0}$ is (and consequently less mixing
with Z) the larger $g_{new}$ should be in order
to preserve the "anomaly" in $R_b$. The following discussion
will be very speculative but helpful to illustrate
a few interesting scenarios.

Let us now imagine there
is a four-fermi coupling of the form: $(g_{s}^2/\Lambda^2)
{\bar {\cal P}} \gamma_{\mu} {\cal P} {\bar f} \gamma^{\mu}
f$, where $f$ denotes any fermion of the first three
generations. The translation of this coupling into
$g_{new}$ is of course highly model-dependent.
The simplest (and most likely easiest to be ruled out) scenario
is to assume that the above coupling comes from the exchange
of some vector boson with point-like coupling to the fermions.
We would then identify $g_{new} \equiv (g_{s}^2/\Lambda^2)
F_V$, meaning that it can be described by the wave function
at the origin. If $g_{s}^2/ 4 \pi =1$ (a strong coupling scenario),
we get $\Lambda = 163-104$ GeV for $M_{V_0} = 95.5-104.5$ GeV,
while if $g_{s}^2/ 4 \pi =2.5$, we would get
$\Lambda = 259-166$ GeV for the same range. These scales are
"uncomfortably" low. Another scenario is to assume that all
fermions are {\em composite} (for instance they could be
bound states of a scalar field and a fermion field). For
definiteness, let us assume that only the fermionic constituents
carry color. A four-fermi coupling given above would be
diagramatically similar to the quark diagram for meson-meson
scattering except that here we would have a scalar line
instead of one of the two quark lines. It follows that
$g_{new}$ is not necessarily given by the wave function at
the origin. We shall assume that we can
write $g_{new} \equiv (g_{s}^2/\Lambda^2)g_H^2 F_V$ where
$g_H^2$ represents the rescattering of the scalar components.
If $g_S^2/ 4 \pi=g_H^2/ 4 \pi= 1, 2.5$, $\Lambda$ can be found to be
respectively 580-371 GeV and 1.45-1.04 TeV for
$M_{V_0}= 95.5-104.5$ GeV. Here $\Lambda$ would represent
the compositeness scale, which as we have seen could be in
the (low)TeV range. A model which we are currently investigating
is similar in spirit to the Abbott-Farhi model except that
the confining gauge group is not the electroweak
group but a horizontal gauge group which we take to
be $SU(2)_L \otimes SU(2)_R$. In this model, there
remains a residual global horizontal $SU(2)$ with
composite fermions forming a triplet (the three
standard families) and a singlet (the {\em non-sequential}
fourth family) under
that group. A full discussion of the model is beyond the
scope of this paper.

We have presented a simple scenario to explain the "anomaly"
in $R_b$ and, as a consequence, we have made a number of
predictions including the presence of a new, {\em non-
sequential} fourth family whose masses could be all below
$M_W$, an exciting prospect for near-future discoveries.

This work was supported in part by the U.S. Department of Energy
under grant No. DE-A505-89ER40518.




%
%
\begin{figure}
\caption{The allowed regions in the $R_{e} ( \equiv
\Gamma (Z \rightarrow had)/
\Gamma (Z \rightarrow e{\bar e}\: or\: \mu{\bar \mu})) - M_{V_0}$ plane.
The theoretical predictions lie above the curves labeled by 300 and 700
for two different values of $m_H$. The vertical lines with similar labels
represent the regions (to the right) allowed by $\Gamma_Z$. The experimentally
allowed regions lie between the two horizontal lines labeled by $R_{e,exp}$.
The intersections of these lines represent the final allowed regions.}
\end{figure}

%
%

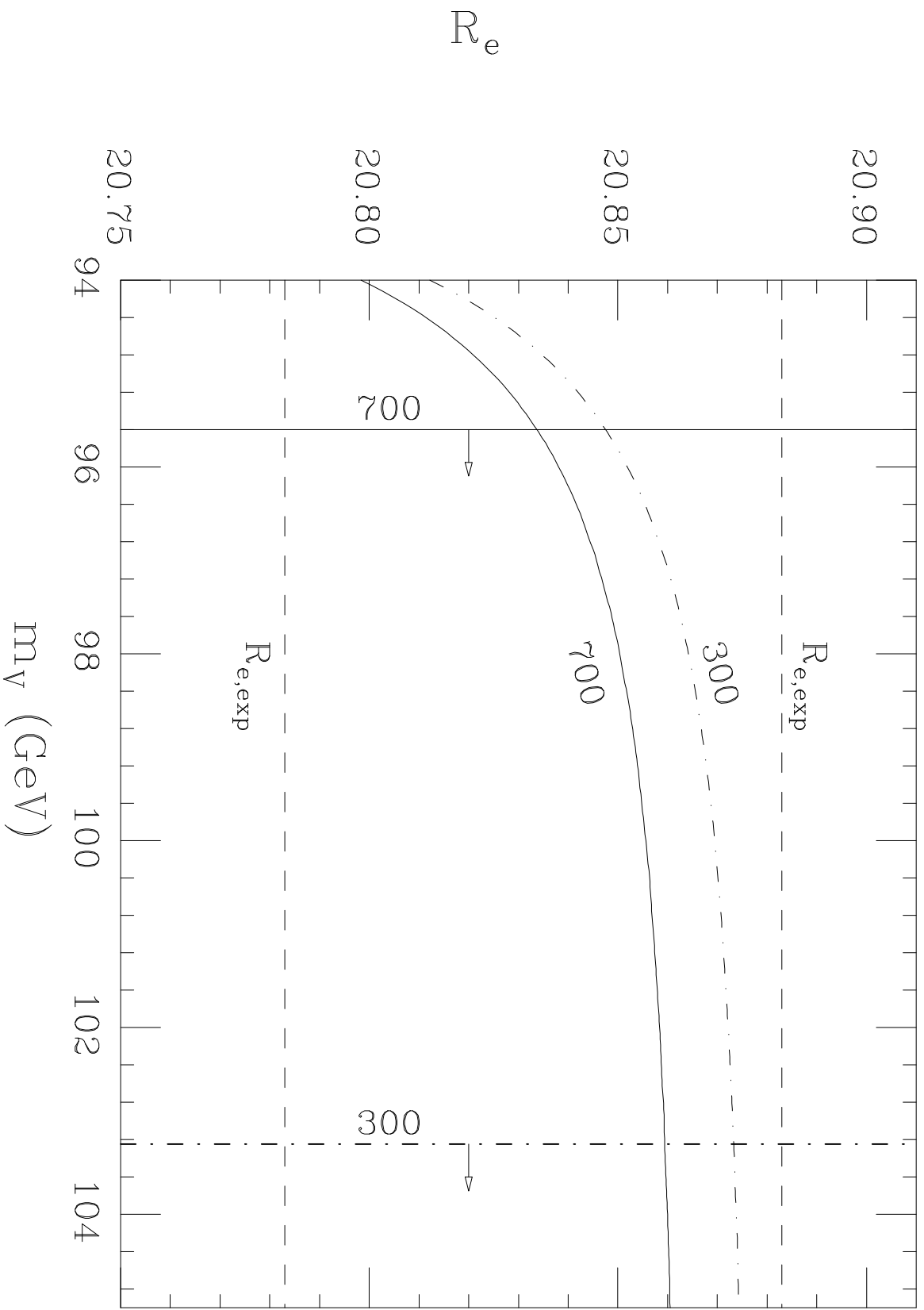
\end{document}